\documentclass[12pt]{article}

\usepackage{graphicx}

\def\be{\begin{equation}}
\def\ee{\end{equation}}

\begin{document}

\title{Impedance of the free surface of liquid electrolytes}

\author{V.~Dashkovsky$^1$ and I.~Chikina$^2$\\
$^1$Institute of Solid State Physics of Russian Academy of
Sciences,\\ 142432, Chernogolovka, Moscow distr., Russia\\
$^2$IRAMIS, LIONS, UMR SIS2M 3299 CEA-CNRS, CEA-Saclay,\\ F-91191
Gif-sur-Yvette Cedex, France}

\date{}

\maketitle

\begin{abstract}
A possibility for the observation of so-called structure
resonances (SR) in electrolytes arising due to relative motion of
the cluster charged nucleus and its solvation shell is
demonstrated. The discussed method considers the resonant
contribution of the SR to the frequency dependence of the
reflection (transmission) coefficient of the electromagnetic wave
interacting with the free electrolyte surface. Of special interest
is the observation of SR for multiply charged particles in
electrolyte providing direct information on the charge of single
cluster. Also important are other not so prominent details of the
wave interaction with mobile charged clusters in electrolyte
related to the formation and complicated nature of the frequency
dependence of the charged cluster associated mass.
\end{abstract}

\section{Introduction}
\label{intro}

The term ``surface impedance'' usually refers to a complex-valued
quantity whose knowledge allows one to calculate the
electromagnetic wave reflection coefficient $R$ from the free
surface of a conductor (e.g., liquid) as well as the absorbtion
and transmission coefficients $A$ and $D$, respectively.
Investigation of these properties of a free surface possesses a
number of advantages deserving special attention. First of all,
field approach minimizes the effects of metal electrodes (which
are normally always present and introduce additional hydrolysis
complications into kinetic problems). Further appealing point
(related to non-symmetric electrolytes) is associated with the
spatial separation multiply charged colloid particles and
compensating ions in the field of image forces. These forces
pushing charged particles from the free surface inside the liquid
are known \cite{1} to be quadratic in the particle charge and,
therefore, affect the equilibrium position of oppositely charged
particles relative to the free surface in different ways. The
arising double layer which has not yet been studied in detail (as
it was already done, e.g. for solutions of charged colloids in
gravity field \cite{2,3,4}) allows to assume that the motion of
separated colloids and corresponding compensating ions along the
liquid surface is to a large extent free from the correlation
phenomena complicating the linear dependence of electrolyte
conductivity on the donor density \cite{5}. Hence, the well-known
impedance formalism  \cite{1, 6} can be without any essential
limitations transferred to liquid non-symmetric conducting media.

Bearing in mind the above arguments, discussed in the present
paper is the impedance approach to the resonance phenomena of
mainly structural origin for non-symmetric electrolytes. The first
part considers the origin of the structure resonances on the whole
and the details relevant to multiply charged clusters. The results
expected here include the possibility of finding the charge of
individual colloid cluster which is one of the most important its
characteristics. The general formalism is also outlined together
with the details of the impedance formalism. The concluding part
of the paper contains results of calculations illustrating the
details of the impedance behavior in a wide frequency range.

\section{Frequency intervals}

Dynamics of charged clusters in liquid electrolytes involves
various frequency-dependent effects in a wide range of frequencies
$\omega$. To describe the details of the electrolyte surface
impedance in various frequency ranges we start with a brief
outline of relevant physics.

We start with the general equations of motion for a charged
cluster in liquid medium in the presence of external electric
field oscillating with frequency $\omega$:
$$
-m_i\omega^2 x=-k_s(x-x_1) +eE_0,  \eqno(1)
$$
$$
 -M\omega^2 x_1=-k_s(x_1-x)+F_{hyd}\left(  \omega, V_1\right)\quad
V_1=-i\omega x_1, . \eqno(2)
$$
Here $m_i$ is the charge bare mass (for simplicity, the charges of
opposite signs are assumed to be identical in all the other
parameters, which is not anyway an essential assumption), $x(t)$
is its oscillating coordinate, $M=4\pi R_i^3\rho_s/3$, $\rho_s
>\rho_l$ is the effective mass of the neutral shell calculated
for the case of a solidified sphere ($\rho_s$ and $\rho_l$ being
the density of solid and liquid, respectively), $x_1(t)$ is the
shell oscillating position, $F_{hyd}\left( \omega, V_1\right)$ is
the hydrodynamic force containing both the viscosity effects and
the cluster frequency dependent associated mass,
$eE_{\parallel}(t) = eE_0\exp{(-i\omega t)}$ is the local external
field of the incident electromagnetic wave. According to (1) the
field $eE(t)$ acts on the particle with mass $m_i$. As to the
neutral part of the cluster with mass $M$, it is driven through
the elastic coupling $k$ with the bare charge.

Equations (1)-(2) allow to roughly identify three interesting
frequency ranges
$$
\omega_s > \omega_M > \omega_{\eta}. \eqno(3)
$$
The first interval, $\omega \sim \omega_s$, contains the so-called
structure resonances for charged clusters in electrolyte. The
frequency range $\omega \sim \omega_M$ is indicated to highlight
the mechanism of formation of the ideal associated mass. Finally,
the third range $\omega \sim\omega_{\eta}$ reveals interesting
details in the behavior of the associated mass of viscous origin.

Following Ref. \cite{7}, we use the term ``structure resonance''
(SR) to identify the resonances occurring in the excitation of the
relative motion between the cluster bare charge and its neutral
shell. According to Eqs. (1)-(2), the structure resonance position
$\omega_s$ is given by the formulae
$$
\tilde\omega^2_s=\frac{1+\gamma}{\gamma}, \quad
\tilde\omega_s=\omega_s/\sqrt{k_s/m_i},\quad \gamma=M/m_i \eqno(4)
$$
provided that the contribution of $F_{hyd}\left(  \omega,
V_1\right)$ into dynamics defined by Eqs. (1)-(2) is relatively
small. Under these conditions the quantity $\tilde\omega^2_s$
varies from unity for $\gamma \gg 1$ up to two for $\gamma \simeq
1$, reproducing the properties of efficient mass well known from
dynamics of a pair of coupled particles. We call the frequency (4)
the dipole frequency since it can be excited by a uniform electric
field. According to Ref. \cite{8}, $F_{hyd}\left( \omega,
V_1\right)$ is small indeed if the cluster compressibility is
characterized by the sound velocity $s_t$ substantially exceeding
that of liquid solvent $s_l$ (or, equivalently, $\rho_s
>\rho_l$).
\begin{figure} %%%%%%Fig.1
\includegraphics*[width=8.50cm]{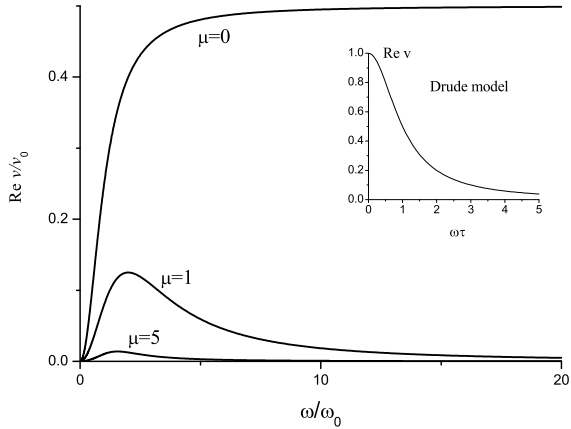} \caption{Re$\,v(\omega)$ (9) as a function
of $\omega$ for different values of the parameter $\mu$. Inset
shows the usual behavior of Re$\,v(\omega)$ for the Drude model
(19).}
\end{figure}
\begin{figure} %%%%%%Fig.2
\includegraphics*[width=8.50cm]{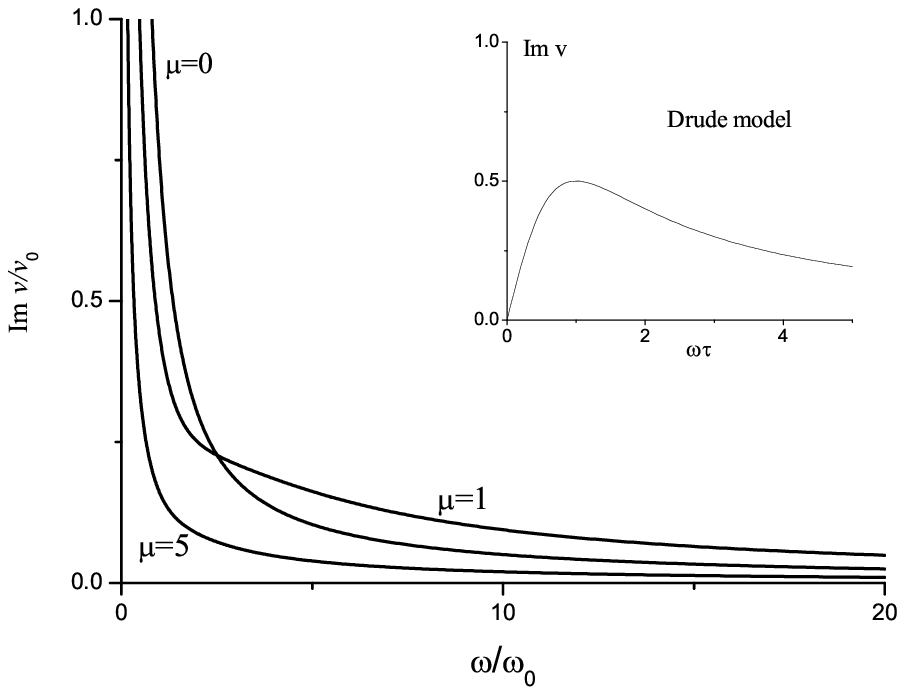} \caption{Im$\,v(\omega)$ (9) as a function
of $\omega$ for different values of the parameter $\mu$. Inset
shows Im$\, v(\omega)$ for the Drude model (19).}
\end{figure}
\begin{figure} %%%%%%Fig.3
\includegraphics*[width=8.50cm]{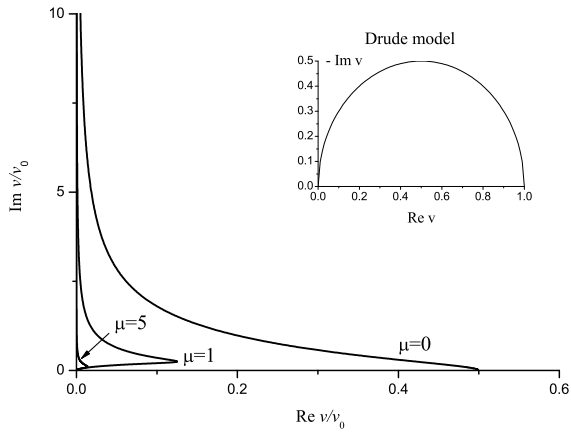}
\caption{Plots of Im$\,v$(Re$\,v$) (9) for the considered model
and the Drude dynamics (inset).}
\end{figure}
\begin{figure} %%%%%%Fig.4
\includegraphics*[width=8.50cm]{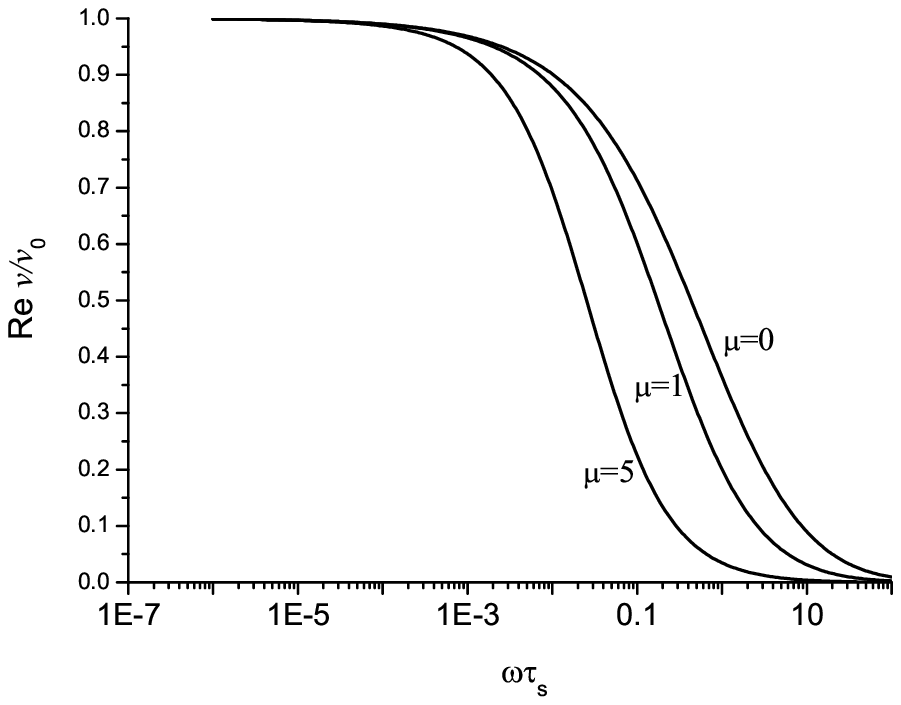} \caption{Re$\,v(\omega)$ (8,11) as a function
of $\omega$ for different values of the parameter $\mu$ for the
Stokes hydrodynamic force.}
\end{figure}
\begin{figure} %%%%%%Fig.5
\includegraphics*[width=8.50cm]{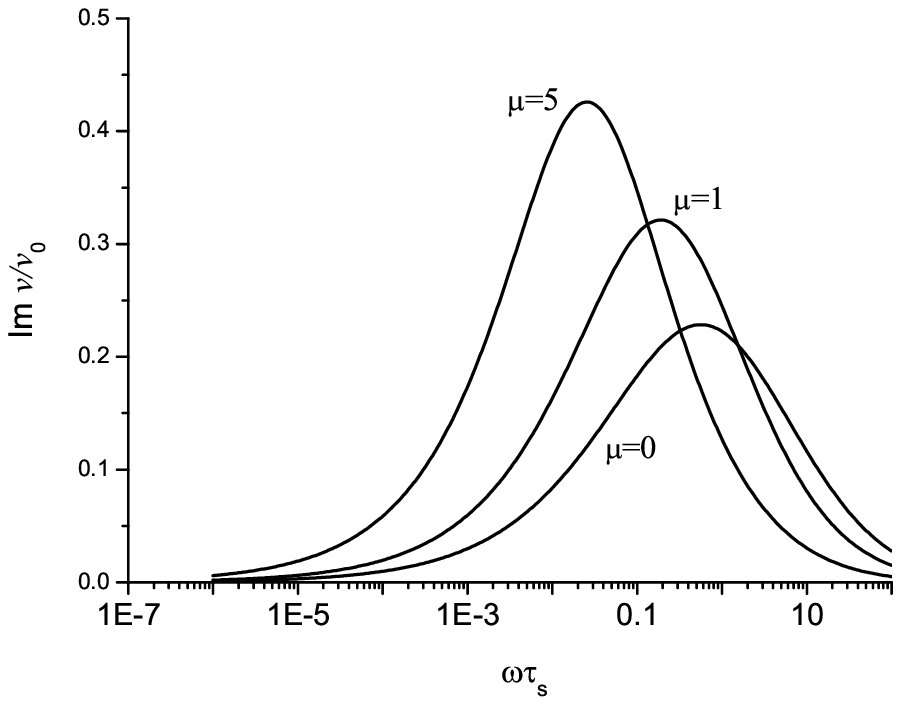} \caption{Im$\,v(\omega)$ as a function
of $\omega$ for different values of the parameter $\mu$ for the
Stokes hydrodynamic force.}
\end{figure}
\begin{figure} %%%%%%Fig.6
\includegraphics*[width=8.50cm]{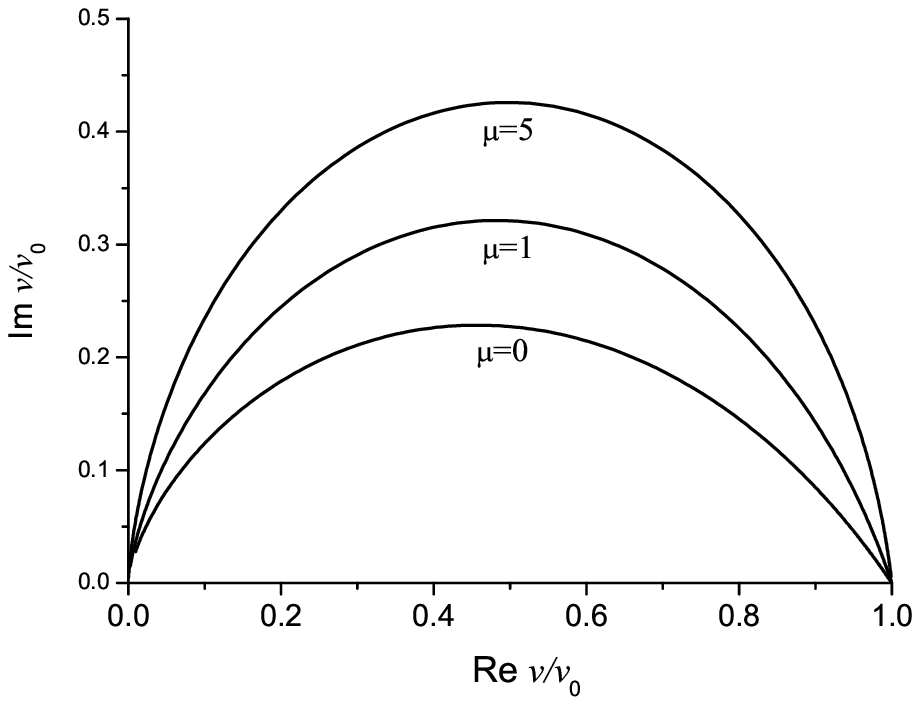}
\caption{Plots of Im$\,v$(Re$\,v)$ for the Stokes hydrodynamic
force.}
\end{figure}

For non-symmetric electrolytes the concept of structure resonance
becomes more complicated. The compensating ion possess the SR of
type (4) with the phenomenological constant $k_s$. As to the
charged colloids, their structure resonance (referred to as
$\Omega_s$ below) can be described in more detail. Here the SR
involves a sort of plasma oscillations in the system of $N$ mobile
charges of mass $m_c$ sliding (presumably) over the surface of the
sphere with radius $R_c$. Simple calculations reveal that the
dipole mode $\Omega_s$ of these oscillations has the scale
$$
\Omega_s^2\simeq \frac{8\pi e^2}{3\epsilon m_c R_c}n_s, \quad n_s=
N/(4\pi R_c^2).\eqno(5)
$$

Here $\epsilon$ is the solvent dielectric constant.

Frequency $\Omega_s$ (5) contains three interesting poorly known
characteristics of the complex: mass $m_c$, effective radius
$R_c$, and total number of charges localized on an individual
colloid particle. Therefore detection of the resonance frequency
$\Omega_s$ allowing to estimate these parameters would be very
useful.

The next two frequency ranges defined by classification (3) can be
considered by employing in Eqs. (1) and (2) suitable asymptotic
expressions for $F_{hyd}\left( \omega, V_1\right)$. For example,
for $\omega \sim \omega_M$ \cite{8}
$$
{\bf F_{hyd}}(\omega)\to {\bf F_{id}}(\omega)=\frac{4}{3}\pi
R_i^3\rho_l\omega{\bf V}
\frac{[-q^3R_i^3+i(2+q^2R_i^2)]}{(4+q^4R_i^4)},\quad \omega\simeq
sq ,\eqno(6)
$$
where $s$ is the sound velocity, $q$ is the wave number (the
dispersion $\omega(q)$ is assumed to be linear), ${\bf V}$ is the
velocity amplitude of the cluster as a whole.

The corresponding associated mass is then
$$
M^{*}(\omega)\simeq{\bf F_{id}}(\omega)/(-i\omega{\bf V}).
\eqno(7)
$$
In the limit $q^2R_i^2\ll 1$ the general definition (7) reduces to
$M^*_0=2\pi R_i^3\rho_l/3$. On the contrary, for $q^2R_i^2\ge 1$
the associated mass $M^{*}(\omega)$ begins to demonstrate a
noticeable frequency dependence. According to Eq. (6), in both
limits oscillations of a sphere in ideal liquid are accompanied by
energy dissipation due to emission of sound waves.

According to the classification (3) and comments to Eq. (6), the
SR position and the range of noticeable mass dispersion are
separated to the extent that the sound velocity in solid exceeds
that in liquid. This circumstance can be used to simplify the
problem of finding the behavior of $M^*(\omega)$ by neglecting the
effects of structure resonances in the frequency range $\omega
\sim \omega_M$. This is formally possible in the limit $k \to
\infty$. In that case the set of equations (1), (2), and (6)
reduces to a single equation
$$
 -(m_i + M)\omega^2 x_1=F_{id}\left(  \omega, V\right)+eE_0,\quad
V=i\omega x_1, \eqno(8)
$$
where $F_{id}\left(  \omega, V\right)$ is taken from Eq.(6) and
$\rho_s> \rho_l$ is the efficient density of the neutral cluster
part, or, in dimensionless variables (frequency $\omega$
normalized $\omega_0 = s_l/R_i$ and velocity $v$ normalized to
$v_0=eE_0/(M^*_0\omega_0)$),
$$
 i\mu\omega v=f_{id}(p, v) + 1,
 \eqno(9)
 $$
 $$
 \mu=(m_i + M)/M^*_0, \quad f_{id}(p, v) = 2\omega v \frac{[-p^3+i(2+p^2)]}{(4+p^4)},
 \quad p=q R_i.
$$

Equations (8,9) allow finding the real and imaginary parts of the
velocity
$$
    {\rm Re} v = \frac{2+(qR)^4/2}{\omega}
                \frac{(qR)^3}{(qR)^6 + (2 + (qR)^2 + \mu(2+(qR)^4/2))^2}; \eqno(10)
$$
$$
     {\rm Im} v = \frac{2+(qR)^4/2}{\omega}
                \frac{2+ (qR)^2 + \mu(2+(qR)^4/2)}{(qR)^6 + (2 + (qR)^2 + \mu(2+(qR)^4/2))^2}.
     \eqno(11)
$$

Thus, we have derived all the quantities required to calculate the
electrolyte impedance in the frequency range of $\omega \sim
\omega_M$.

The frequency range $\omega \sim \omega_{\eta} <\omega_M$ is
interesting because of unusual viscosity effects on the associated
mass. Here the ideal associated mass is already fully developed
and one can employ the concept of liquid viscosity $\eta$. In this
case the solution of the Navier-Stokes equation yields the
following frequency representation for the force ${\bf
F_{hyd}}(\omega)\to F_s\left(\omega, V\right)$ \cite{8}:
$$
F_s\left(  \omega, V\right) =6\pi\eta R_{c}\left( 1+\frac{R_{s}}
{\delta\left(  \omega\right)  }\right) V\left( \omega\right)
 +3\pi
R_{s}^{2}\sqrt{\frac{2\eta\rho}{\omega}}\left(
1+\frac{2R_{s}}{9\delta\left(  \omega\right)  }\right)  i\omega
V\left( \omega\right)  ,
 \eqno(12)
$$
Here $V$ is the sphere velocity as a whole and $\delta\left(
\omega\right) =\left( 2\eta/\rho_l\omega\right) ^{1/2}$ is the
so-called dynamic penetration depth.

Equations of motion now acquire the form of (8) with $F_{id}\left(
\omega, V_1\right)$ from (6) replaced with $F_s\left( \omega,
V\right)$ (12). The coefficient at the imaginary part in Eq. (11)
measuring inertia contribution (i.e., the associated mass) to the
general expression for the force $F_s$ proves to be divergent as
$\omega^{-1/2}$ with decreasing frequency. This fact deserves
special attention in itself.

\section{Impedance details}
Going back to the details of the impedance, we start with a few
general relations. If Ohm's law in the medium is written as
$$
\bf{j}=\alpha\bf{\dot E}+\sigma\bf{E} \eqno(13)
$$
where $\alpha$ and $\sigma$ are real constants, then its
refraction index $n$ and absorption index $k$ can be expressed
through $\alpha$ and $\sigma$ in the following way \cite{1},
\cite{6}:
$$
(1+4\pi\alpha)=n^2-k^2, \quad \sigma=nk\omega/2\pi \eqno(14)
$$

In its turn, the reflection coefficient $R$ for the electrolyte
surface is
$$
R=\frac{(n-1)^2+k^2}{(n+1)^2+k^2} \eqno(15)
$$
In all the preceding formulas (13--15) the electrolyte magnetic
permeability was assumed to be equal to unit.

 Current (13) in the medium is due to the motion of opposite
charges preserving local neutrality. In three-dimensional problems
local neutrality holds not only in the equilibrium, but also in
the linear regime (13) under the applied vortex fields (in 3D
systems the charge build up is associated with the violation of
the condition $div E=0$; in the transverse wave this condition is
assumed to be satisfied). Thus, the problem of the wave
interaction with semi-infinite electrolyte reduces to finding the
constants $\alpha$ and $\sigma$ appearing in Eq. (12) from
single-particle equations (1, 2).
%and related definitions (8, 9).

The simplest variety of the interaction of the charged complex
(1), (2) with the electromagnetic wave arises if one assumes that
the shell motion is completely suppressed (large values of the
mass $M$ or, which is more realistic, finite values of the solvent
viscosity $\eta$, damping through the force $F_{hyd}\left( \omega,
V_1\right)$ in Eqs. (1), (2) the free shell motion). In that case
only the bare charge keeps its mobility so that the general
picture proves to be essentially dielectric, and the problem (1),
(2) for the oscillatory motion of a separate charge is reduced to
the well known problem of dielectric constant dispersion in a
homogeneous system of oscillators distributed with the spatial
density $n_i$. The effective high-frequency conductivity in that
case is
$$
\tilde\sigma=\frac{n_i
e^2}{m_i}\frac{\omega\exp{[i(0.5\pi-\varphi)]}}
{\sqrt{(\omega_s^2-\omega^2)^2+\gamma^2\omega^2}}=i\omega\alpha+\sigma.
\eqno(16)
$$
Here $\omega_s$ is the SR position, $\gamma$ is the decay of the
charge oscillations within the cluster, $\varphi$ is the current
delay phase with respect to voltage. In accordance with Eq. (12),
the refraction index can then be obtained as
$$
n^2-k^2=1+\frac{4\pi n_i e^2}{m_i}\frac{(\omega_s^2-\omega^2)}
{(\omega_s^2-\omega^2)^2+\gamma^2\omega^2}\eqno(17)
$$
$$
2nk=\frac{4\pi n_i e^2}{m_i}\frac{\gamma\omega}
{(\omega_s^2-\omega^2)^2+\gamma^2\omega^2}.\eqno(18)
$$

Assuming $(\omega_s^2-\omega^2)^2\gg\gamma^2\omega^2$, one can
take $k\to 0$, where $k$ is from (14). Then
$$
n^2\simeq 1+\frac{4\pi n_i e^2}{m_i}\frac{1}
{(\omega_s^2-\omega^2)},\eqno(19)
$$
which demonstrates the well known fact of the appearance of
refraction index frequency dispersion in the ensemble of
oscillators.

When dealing with the frequency $\Omega_s$ (5), it is reasonable
not only to calculate the resonance position (19) but also find
the imaginary part of $k$ defined by Eq. (18). This is useful for
extracting from experimental data of both $N$ and the effective
mass $m_c$.  In that case the radius $R_c$ should be additionally
obtained from low frequency measurements of the colloid mobility.

In the other frequency ranges ($\omega_M$ and $\omega_{\eta}$)
indicated in Eq. (3) the finite cluster mass $M$ and solvent
viscosity $\eta$ the shell mobility cannot be neglected. One
should solve the problem of cluster behavior in the external field
more accurately, mainly in the sense of actual properties of the
force $F_{hyd}\left( \omega, V_1\right)$. This approach (which was
described in detail in the preceding part of the paper) allows one
to immediately formulate the final results for frequency ranges
defined by Eq. (3).

For convenience, in the frequency ranges $\omega \sim \omega_M$
and $\omega \sim \omega_{\eta}$ we consider the most significant
and graphical (in our opinion) information concerning the
structure of the corresponding hodographs expressing direct
relations between ${\rm Re} V$ and ${\rm Im} V$ against the
background of respective plots for the case of Drude dynamics,
$$
m(-i\omega+\tau^{-1})v=eE_0. \eqno(20)
$$
Here $m$ is some scalar mass and $\tau$ is the typical frequency
independent relaxation time.

Frequency range $\omega \sim \omega_M$. Comparative plots for the
frequency dependence ${\rm Re} v$ and ${\rm Im} v$ calculated for
different values of the parameter $\mu = (m_i + M)/M^*_0$  (see
Eq. 8) as well as the corresponding hodograph are shown in Figs.
1--3. This parameter reveals the role of the bare mass $M$ in the
considered problem. In particular, the values $m_i=M=0$,  models
the limit of zero bare mass and zero solidified neutral shell mass
occurring for single-electron bubbles in liquid helium.

Frequency range $\omega \le \omega_{\eta} <\omega_M$. Similar
plots obtained by employing Eq. (8) with the hydrodynamical force
defined by Eq.(11) are presented in Figs. 4--6.

Summary. Suggested is a resonance technique for the study of the
structure of charged clusters in various electrolytes. The
resonances, which were called structure resonances, arise in the
course of excitation of relative motion of the bare charge and the
neutral cluster shell, and contain interesting information on the
strength of their elastic coupling.  Within the framework of the
developed formalism the problem of formation of ``ideal''
associated mass of oscillating sphere is discussed which is
directly related to the behavior of structure resonances.

The work was supported by the RFBR grant No. 09-02-00894a and
Program of the Presidium of RAS ``Quantum Physics of Condensed
Matter''.

\end{document}